\documentclass[twocolumn,showpacs,preprintnumbers,amsmath,amssymb,prb,superscriptaddress]{revtex4}

\usepackage{graphicx}

\begin{document}

\title{Hamiltonian approach to the charge transfer statistics of Kondo quantum dots contacted by a normal metal and a superconductor}
\author{H. Soller$^1$ and A. Komnik}
\affiliation{Institut f\"ur Theoretische Physik,
Ruprecht-Karls-Universit\"at Heidelberg,\\
 Philosophenweg 19, D-69120 Heidelberg, Germany}
\date{\today}

\begin{abstract}
We analyze the full counting statistics (FCS) of quantum dots in the Kondo regime contacted by normal and superconducting leads or an STM tip. To describe the Kondo resonance we use an effective model for the quantum dot in the Kondo regime in combination with the FCS for tunneling contacts calculated using the Hamiltonian approach. We show that the situation of weak coupling to the superconducting electrode in the case of superconductor gap being larger than the Kondo resonance width can be easily handled and verify the method by comparing our theoretical predictions to experimental data. This allows us to make predictions for the noise and cross-correlation in setups involving a superconductor. We find that a positive cross-correlation is possible in the case of a superconductor and two normal leads contacted via two quantum dots in the Kondo regime.
\end{abstract}


\maketitle

\section{Introduction}
Accessing electronic transport in nanoscale systems requires an understanding not only of the current but also its fluctuations. Whereas charge transfer between normal metals proceeds via single electron transfer processes, in normal-superconductor hybrid structures the current transport at energies below the superconducting gap is governed by Andreev reflection.\cite{ajetp} An electron from the normal conductor cannot enter the superconductor at energies below the gap and is therefore retroreflected as a hole leaving a Cooper pair behind.

In recent years it has become possible to not only make nanometer sized junctions \cite{PhysRevB.69.140502,PhysRevLett.93.128303} but also to contact a single impurity via two leads.\cite{goldhaber,Schmid1998182} This configuration and the corresponding Anderson impurity model allows to study the Kondo effect\cite{PhysRevB.83.075107,PhysRevB.81.165106,PhysRevB.79.035320} in many geometries. \cite{Cronenwett24071998,PhysRevLett.70.2601,PhysRevB.63.094515,Braggio2011155,PhysRevLett.93.047002}  Normal-quantum dot-superconductor hybrid structures are of special interest since they represent the main building blocks of Andreev entanglers.\cite{19829377,PhysRevLett.104.026801,2009arXiv0910.5558W,0295-5075-54-2-255,PhysRevLett.74.3305.2}

To obtain a complete knowledge of the charge transfer in these systems it is not sufficient to study the nonlinear $I-V$ and the shot noise alone. It is very useful to possess a complete information about the probability distribution of charge transfer, also referred to as full counting statistics (FCS). It is usually calculated in terms of a counting field $\lambda$-dependent cumulant generating function (CGF) of charge transfer.\cite{levitov-1996-37,nazarov-2003-35} Derivatives of the CGF with respect to the counting field directly generate all irreducible moments of the distribution. Additionally the type of distribution and the actual $\lambda$-dependence allows for a rigorous identification of the elementary processes of charge transfer.\cite{PhysRevB.63.201302,PhysRevLett.98.056603,PhysRevLett.87.197006,PhysRevB.70.115305} Although the experimental access to higher order cumulants is rather difficult it has already been demonstrated in a number of works.\cite{reulet-2003-91,epub12308}

The FCS for quantum point contacts (QPCs) between normal and/or superconducting leads have been considered before.\cite{levitov:4845,PhysRevB.50.3982,fcsons,PhysRevLett.87.197006,PhysRevLett.87.067006,PhysRevB.66.201306} Likewise the FCS for a quantum dot in the Kondo limit contacted by normal drains have been calculated at the Toulouse point.\cite{PhysRevB.75.235105,PhysRevB.73.195301} However, these techniques are hard to generalise to the case of a superconducting drain. Likewise the analytical techniques developed so far for the $I-V$ characteristics of normal-quantum dot-superconductor devices\cite{PhysRevB.63.094515,PhysRevB.77.104525} even in the simplest cases lead to rather cumbersome calculations.

In this work we want to show how results for the FCS of tunneling junctions allow for effective descriptions for the situation of a quantum dot in the Kondo limit that is coupled to a superconductor or an STM tip. We analyse the interplay of the Kondo effect\cite{KOU01} and superconductivity\cite{PhysRevB.63.094515,0957-4484-15-7-056} for situations in which $\Gamma_K \lesssim \Delta_0$, where $\Delta_0$ is the superconductor gap and $\Gamma_K$ is the Kondo resonance width. We also assume a strong asymmetry of the couplings of the quantum dot to the normal electrode and the superconductor/STM tip. While an STM tip represents a special normal contact the asymmetry is generic in the case of a normal-quantum dot-superconductor junction in this regime.\cite{0957-4484-15-7-056} As a first step we derive the FCS for a normal-superconductor junction using the Hamiltonian approach as introduced in [\onlinecite{PhysRevB.54.7366}]. In this approach to quantum transport nanostructures are analysed starting from a microscopic Hamiltonian, which essentially describes transport as tunnelling contrary to the alternative strategies of [\onlinecite{PhysRevB.50.3982,fcsons}] modelling transport by scattering. Our approach allows us to formulate the effective model for a coupled quantum dot. We find that the explicit consideration of Andreev reflection in normal-quantum dot-superconductor junctions improves the fit to the experimental conductance compared to previous studies and leads to a considerable enhancement of the Fano factor for voltages below the gap. We also show how Andreev reflections in a junction with two quantum dots lead to a positive cross correlation of the currents in the normal leads.

\section{Quantum Point Contacts}
The first system we want to consider is a QPC between a normal metal and a superconductor/STM tip as in [\onlinecite{levitov:4845,PhysRevB.50.3982,fcsons}]. It is described by the following Hamiltonian
\begin{eqnarray}
H=H_1 + H_2 + H_T\,. \label{system}
\end{eqnarray}
$H_1$ refers to the normal metal, which is modelled as a noninteracting fermionic continuum at chemical potential $\mu_1$, written in terms of electron field operators $\Psi_{k1\sigma}$. $H_2$ either refers to another normal metal or a superconducting reservoir that is described by the BCS Hamiltonian\cite{PhysRev.108.1175} with the gap $\Delta_0$ of the superconducting terminal
\begin{eqnarray*}
H_2 &=& \sum_{k,\sigma} \epsilon_k \Psi_{k2\sigma}^+ \Psi_{k2\sigma} \nonumber\\
&& + \Delta_0 \sum_k \left(\Psi_{k2\uparrow}^+ \Psi_{-k2\downarrow}^+ + \Psi_{-k2\downarrow} \Psi_{k2\uparrow}\right).
\end{eqnarray*}
As in previous treatments of the problem,\cite{PhysRevB.50.3982} the second lead is kept at $\mu_2 = 0$, so that the applied bias voltage is given by $V= \mu_1 - \mu_2=\mu_1$ (we use units $e = \hbar = k_B = 1$).
Electron hopping between the two electrodes is described by the Hamiltonian\cite{PhysRevLett.8.316}
\begin{eqnarray*}
H_T = \sum_{\sigma} \gamma \left[\Psi_{1\sigma}^+(x=0) \Psi_{2\sigma}(x=0) + h.c.\right],
\end{eqnarray*}
where $\gamma$ is the amplitude of the respective tunneling coupling and the tunneling is assumed to occur at $x=0$ in the coordinate system of the electrodes.

The density of states (DOS) in the vicinity of the Fermi level in the normal electrode $\rho_{01}$ is assumed to be only weakly energy-dependent. For a superconducting drain the DOS is strongly energy-dependent and diverges for energies close to the superconducting gap. From BCS theory\cite{PhysRev.108.1175} it is known to be given by
\begin{eqnarray*}
\rho_2 = \rho_{02}\frac{|\omega|}{\sqrt{\omega^2 - \Delta_0^2}}.
\end{eqnarray*}
In this work we want to use the Hamiltonian approach, meaning the complete resummation of perturbation theory in the tunneling coupling. Instead of coupling the counting field directly to the current operator as in [\onlinecite{PhysRevB.50.3982,fcsons}], $\lambda$ enters the tunneling part of the Hamiltonian according to the prescription\cite{PhysRevB.70.115305,PhysRevB.73.195301}
\begin{eqnarray}
T^\lambda = \sum_{\sigma} \gamma \left[ e^{-i \lambda/2} \Psi_{1\sigma}^+(x=0) \Psi_{2\sigma}(x=0) + \mbox{h.c.}\right] \, \label{tl}.
\end{eqnarray}
The counting field is explicitly time-dependent: it is zero outside the measurement interval $0 < t < \tau$ and carries different signs on the forward/backward Keldysh branches.
The CGF is given by the expectation value: $\ln \chi(\lambda) = \ln \langle e^{i \lambda Q}\rangle$ that allows to compute the current, noise and also higher statistical moments that may become experimentally observable in these systems. The different cumulants of the distribution are obtained by successive differentiation with respect to the counting field. The CGF for long measurement times $\tau$ can be calculated analytically using a Green's function method\cite{nazarov,levitov-1996-37} specifically designed for studying quantum impurity problems.\cite{PhysRevB.73.195301,bagrets-2006-54,maier}

For the FCS calculation we follow the approach outlined in [\onlinecite{PhysRevB.73.195301}]: for long measurement times the CGF may be written using an adiabatic potential\cite{PhysRevLett.26.1030} (that by construction does not depend on time) as
\begin{eqnarray*}
\ln \chi(\lambda) = -i \int_0^\tau dt\, U(\lambda) = -i \tau U(\lambda) \,.
\end{eqnarray*}
In turn the adiabatic potential is related to the counting field derivative of $T^{\lambda}$ in Eq. (\ref{tl})
\begin{eqnarray}
&& \frac{\partial U}{\partial \lambda} = \left \langle \frac{\partial T^\lambda}{\partial \lambda} \right\rangle_\lambda\nonumber\\
&=& \left\{\frac{-i \gamma}{2} \sum_\sigma \langle T_{\cal C} \Psi_{1\sigma}^+(x=0) \Psi_{2\sigma}(x=0) \rangle_\lambda + \mbox{h.c.}\right\}, \label{adiabatic}
\end{eqnarray}
where $\langle \cdot \rangle_\lambda$ is defined as $\langle \cdot \rangle_\lambda := 1/\chi(\lambda) \langle \cdot \rangle$ with $\langle \cdot \rangle$ being the ordinary expectation value with respect to the system's Hamiltonian in Eq. (\ref{system}) with $H_T$ replaced by $T^\lambda$ in Eq. (\ref{tl}). The mixed and $\lambda$-dependent Green's functions in Eq. (\ref{adiabatic}) are calculated exactly by summing up all orders in the tunneling coupling generated by $T^\lambda$. The FCS for the case of a contact between two normal metals is well known\cite{levitov-1996-37} [see also Eq. (\ref{fcsnn})]. In the case of a contact between a normal metal and a superconductor contrary to [\onlinecite{PhysRevB.73.195301}], of course, we have to use superconducting Green's functions for the superconducting side as given e.g. in [\onlinecite{PhysRevB.80.184510}]. In this way we obtain an exact expression for the temperature and energy-dependent CGF that is valid over the whole range of possible parameters. It is given by
\begin{widetext}
\begin{eqnarray*}
\ln \chi_{NS}(\lambda, \tau) &=&  \tau \int \frac{d\omega}{\pi} \left\{\ln \left\{\prod_{\alpha = \pm} \left\{1+ T_e(\omega) \left[n_{1\alpha} (1-n_2) (e^{i \alpha \lambda} -1) + n_2(1-n_{1\alpha}) (e^{-i \alpha \lambda}-1)\right]\right\}\right.\right. \nonumber\\
&& + T_{A2}(\omega) (2n_2-1) \left\{ (2n_2-1) \left[(e^{i \lambda} -1)^2 n_{1-}(1-n_{1+}) - 2 (e^{i \lambda}-1)(e^{- i \lambda}-1) n_{1-} n_{1+} \right. \right.
\end{eqnarray*}
\end{widetext}
\begin{widetext}
\begin{eqnarray}
&& \left. \left. + (e^{-i \lambda} -1)^2 n_{1+} (1-n_{1-})\right] + 2n_2(e^{i \lambda} -1) (e^{-i \lambda} -1) (n_{1+} -1 + n_{1-})\right\}\nonumber\\
&& + T_{BC} (\omega) (2n_2-1) (e^{i \lambda} - e^{-i \lambda})^2 \left\{ (2n_2-1) [n_{1-} e^{i \lambda} + n_{1+} e^{-i \lambda} + \Gamma_{e} n_2 (1-n_2) (e^{i \lambda} - e^{-i \lambda})^2 \right. \nonumber\\
&& \left. \left. - (n_{1+}-1 + n_{1-}) n_2 (e^{i \lambda} + e^{-i \lambda})] - 4n_2 (1-n_2) (n_{1+} -1 +n_{1-})\right\} \theta\left(\frac{|\omega| - \Delta_0}{\Delta_0}\right)\right\} \nonumber\\
&& + \left. \ln \left\{1+ T_A(\omega) \left[n_{1+} (1-n_{1-}) (e^{2i \lambda}-1) + n_{1-}(1-n_{1+}) (e^{-2i\lambda}-1)\right]\right\} \theta\left(\frac{\Delta_0 - |\omega|}{\Delta_0}\right)\right\}, \label{fcssn}
\end{eqnarray}
\end{widetext}
where the effective transmission coefficients are given by
\begin{eqnarray}                         
T_e(\omega) &=& \frac{4\Gamma_{e}}{[(1+ \Gamma_{e})^2 - \Gamma_{A}^2]}, \nonumber\\
T_{A2}(\omega) &=& \frac{4 \Gamma_{A}^2}{[(1+ \Gamma_{e})^2 - \Gamma_{A}^2]^2} = \frac{T_{BC}(\omega)}{\Gamma_e} \; \mbox{and} \nonumber\\
T_A(\omega) &=& \frac{4 \Gamma_{A}^2}{\Gamma_{A}^4+ 2 \Gamma_A^2(1- \Gamma_e^2) + (1+ \Gamma_e^2)^2}. \label{trans}
\end{eqnarray}
The energy-dependent DOS of the superconducting terminal affects the hybridisations $\Gamma_{e} = \Gamma |\omega|/\sqrt{|\omega^2 - \Delta_0^2|}$ and $\Gamma_{A} = \Gamma \Delta_0/\sqrt{|\Delta_0^2 - \omega^2|}$, where $\Gamma = \pi^2 \rho_{01} \rho_{02} \gamma^2/2$. The subscripts $e$ and $A$ refer to the electronic and Cooper pair density of states. $T_e$ refers to the transfer of single electrons and $T_{A2}$, $T_{BC}$ describe the additional noise contributions by Andreev reflection above the gap and branch crossing processes respectively.\cite{PhysRevB.54.7366} $T_A$ is the transmission coefficient of Andreev reflection processes below the gap. The Fermi distribution for the normal/superconducting lead is abbreviated by $n_{1+,2}$. $n_{1-} = 1- n_{1+}(-\omega)$ refers to hole-like contributions.

The expression in Eq. (\ref{fcssn}) is the first result of this work. For $\Delta_0 \rightarrow 0$ only the first line of the expression remains which is the sum of electrons and holes of the the expression for the CGF of a normal tunnel contact derived in [\onlinecite{levitov-1996-37}]. Consequently the dominating charge transfer events above the gap are single electron transfers. Below the gap only terms proportional to $e^{2i\lambda}$ occur referring to double electron transfers in Andreev reflection processes.\footnote{Needless to say, this is consistent with the properties of the scattering matrix, which has two energy regimes.\cite{PhysRevB.50.3982,schwab-1999-59}} Since the result in Eq. (\ref{fcssn}) is an exact expression this interpretation is unambiguous contrary to perturbative results where the analysis is often not trivial.\cite{PhysRevB.72.235328,PhysRevB.73.195301} Using the scattering approach the FCS of a similar system has been studied before.\cite{PhysRevB.50.3982,fcsons} Because a direct comparison of the CGF is quite difficult we start by comparing the lowest cumulants and find excellent agreement.\\

In particular, we compute the current corresponding to this CGF via
\begin{eqnarray}
I_{NS} = - \frac{i}{\tau} \frac{d}{d\lambda} \ln \chi_{NS}(\lambda, \tau)|_{\lambda = 0}.
\end{eqnarray}
Considering the differential conductance $d I_{NS}/dV$ for low voltages and $T=0$ we obtain the result previously obtained in [\onlinecite{beenakker}]. Likewise the equivalence of the widely used BTK model\cite{PhysRevB.25.4515} and the Hamiltonian approach used here has been demonstrated in [\onlinecite{PhysRevB.54.7366}]. We also reproduce exactly the results for the nonlinear current voltage characteristics in [\onlinecite{PhysRevB.54.7366}]. To further demonstrate the validity of the Hamiltonian approach we also performed a comparison to the experimental results for Al/Cu contacts with high transparency presented in [\onlinecite{PhysRevB.69.140502}] and obtained a very good agreement.\\

We also compared our result directly to the FCS obtained in [\onlinecite{PhysRevB.50.3982,fcsons}]. In [\onlinecite{PhysRevB.50.3982}] it was argued that one should use the transmission coefficients from the Blonder-Tinkham-Klapwijk (BTK) model\cite{PhysRevB.25.4515} and in Ref.~[\onlinecite{fcsons}] circuit theory in combination with the BCS Green's functions reproduces this result. In [\onlinecite{PhysRevB.54.7366}] it was shown how to obtain the BTK transmission coefficients from the result using the Hamiltonian approach. Using this mapping on Eq. (\ref{fcssn}) we recover the results obtained in [\onlinecite{PhysRevB.50.3982,fcsons}].
From Eq.~(\ref{fcssn}) it is evident that the FCS has two regimes: for $|V| \gg \Delta_0$ only single-electron processes constitute the current while for $|V| \ll \Delta_0$ only Andreev processes contribute.

\section{Quantum dots in the Kondo regime contacted by superconductors or STM tips}

\begin{figure}[ht]
\includegraphics[width=7cm]{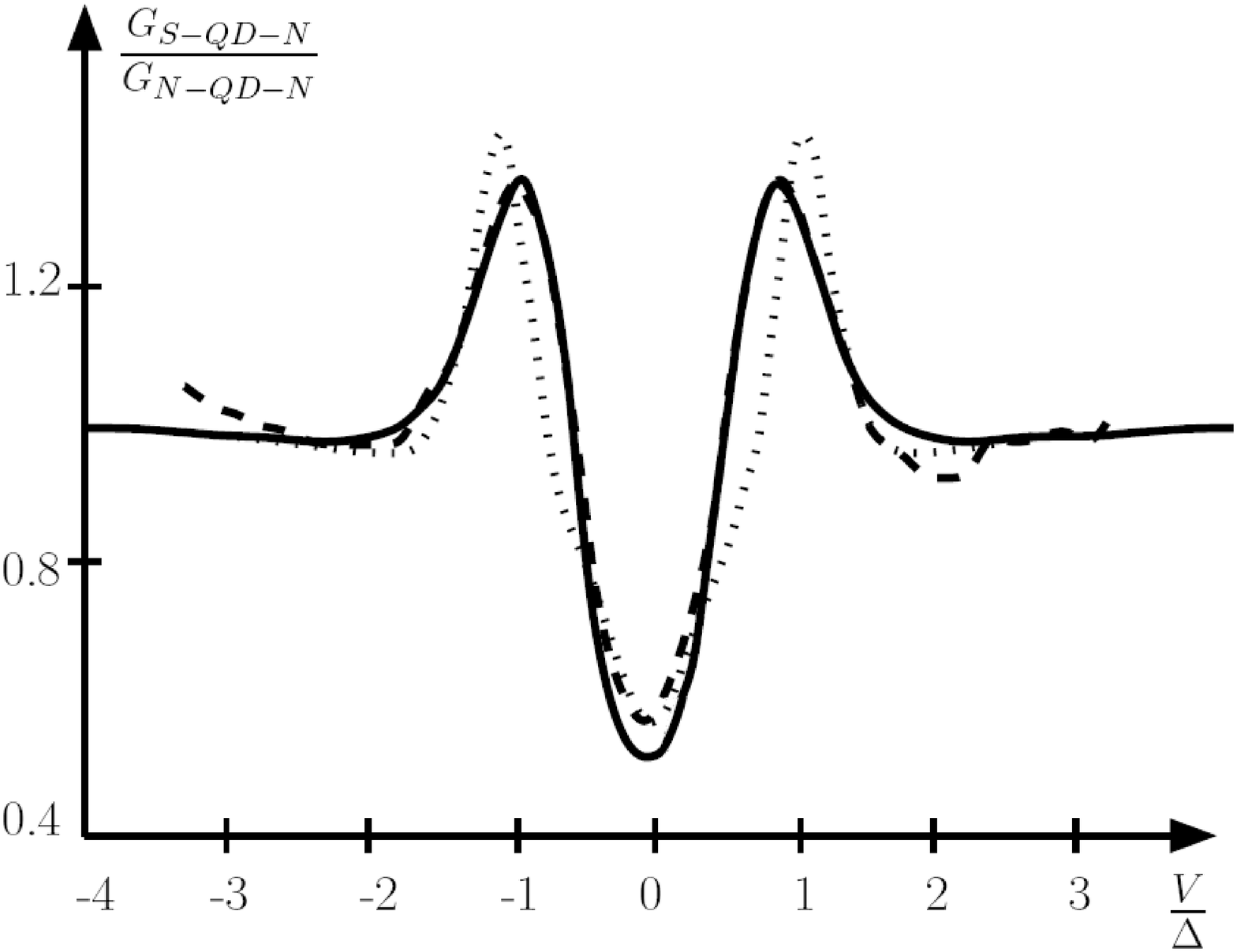}
\caption{Theoretical prediction for the differential conductance $G_{S-QD-N}$ divided by the normal state value $G_{N-QD-N}$ shown as the solid curve. The parameters taken from [\onlinecite{0957-4484-15-7-056}] include $\Gamma_K = 0.7\Delta$, $\Gamma =0.05$, $\Gamma_D = 0.053\Delta$, $T_b = 0.375$ and the width of the Lorentzian for the conductance in the KNQN case of $0.6\Delta$. To obtain the peaks at the right position for our fit we need to employ a slightly lower superconducting gap compared to the fit parameter $\Delta$ in the semiconductor model\cite{tinkham} (dotted curve) taken also from [\onlinecite{0957-4484-15-7-056}] $\Delta_0 = 0.7 \Delta$. The result can be compared to the experimental data shown as the dashed curve.\cite{0957-4484-15-7-056}}
\label{fig4}
\end{figure}

The Hamiltonian approach can be applied in more complex geometries, e.~g. for the resonant level system, which is the simplest quantum dot. Including Coulomb interaction on the quantum dot we observe a Kondo resonance when the number of electrons on the quantum dot is odd so that it obtains a localized magnetic moment of spin $1/2$. At energy scales below the Kondo temperature $T_K$ the dot spin hybridizes with the lead spin density and forms a singlet state. At this strong-coupling fixed point a perfectly transmitting channel opens up and the Kondo effect can be described as a pure resonant level system as far as the electronic transport is concerned.\cite{aleiner} For weak coupling of the second lead to the quantum dot one can see from the calculations in [\onlinecite{PhysRevB.73.195301,soller}] that a resonant level leads to a renormalization of the effective transmission coefficient for a tunnel contact with the density of states of a resonant level. This situation emerges in the case of a quantum dot in the Kondo regime contacted by a normal lead and an STM tip (KNQN) or by a superconductor provided $\Gamma_K \lesssim \Delta_0$ holds\cite{0957-4484-15-7-056} (KNQS). In the opposite case $\Gamma_K > \Delta_0$ the Kondo resonance also strongly couples to the quasiparticles on the superconducting side leaving the current voltage characteristics almost identical to those obtained for normal conducting systems.\cite{0957-4484-15-7-056} If only one drain strongly couples to the quantum dot the coupling to the second lead can be described by an effective tunneling transmission. We can thus employ the transmission coefficients for the QPCs multiplied with the Kondo density of states $\Gamma_K^2/\left[(\omega - V)^2 + \Gamma_K^2\right]$.\cite{0957-4484-15-7-056}\\
If the Kondo impurity is contacted by two normal drains with strongly asymmetric coupling its transmission can be effectively written as 
\begin{eqnarray}
T_{enK} = \frac{4\Gamma_{NN}}{(1+ \Gamma_{NN})^2} \frac{\Gamma_K^2}{\left[(\omega - V)^2 + \Gamma_K^2\right]},
\end{eqnarray}
where $T_{en} = 4 \Gamma_{NN}/(1+ \Gamma_{NN})^2$, where $\Gamma_{NN} = \pi^2 \rho_{01} \rho_{02} \gamma^2$ is the transmission coefficient of the ordinary QPC. The factor $\Gamma_K^2/[(\omega - V)^2 + \Gamma_K^2]$ accounts for the Kondo resonance.\\
Additionally we have to include the density of states outside the Kondo peak given e.g. by the Hubbard subbands at $\pm U/2$, where $U$ refers to the strength of Coulomb repulsion on the quantum dot. In a first approximation this can be modelled by a constant background conductance $T_b$ as in [\onlinecite{0957-4484-15-7-056}]. The FCS of these additional processes can be described by the standard Levitov-Lesovik formula\cite{levitov-1996-37}
\begin{eqnarray}
&& \ln \chi_{b}(\lambda, \tau) = 2\tau \int \frac{d\omega}{2\pi} \ln \{1+ T_{b}\nonumber\\
&& \times [(e^{i \lambda}-1)n_{1+}(1-n_2) + (e^{-i \lambda}-1) n_2 (1-n_{1+})]\}. \label{fcsnn}
\end{eqnarray}
The total CGF is given by $\ln \chi_{KN} = \ln \chi_{KNQN} + \ln \chi_{b}$, where $\chi_{KNQN}$ can be derived from $\chi_{b}$ by replacing $T_{b}$ by $T_{enK}$.\\
As expected, for the differential conductance we observe the typical Lorentzian shape that is given by the form of the Kondo resonance. However, in this case there is no energy scale but the width of the Kondo resonance.\\
This is different in the case of a superconducting lead with its characteristic energy gap. E.g. for two superconducting terminals the Kondo effect is suppressed if the superconducting gap is larger than $T_K$.\cite{PhysRevLett.89.256801} A similar effect occurs in the regime $\Gamma_K \lesssim \Delta_0$ considered here in a KNQS junction since only electrons on the normal side strongly couple to the emerging Kondo resonance. One therefore obtains a strong asymmetry of the Kondo coupling between the normal and the superconducting drain to the dot.\cite{0957-4484-15-7-056} Consequently, one observes a strong suppression of the total conductance while side peaks at the superconducting gap appear. The assembly of superconductor-quantum dot-hybrid structures with good coupling to the quantum dot makes $\Gamma_K \lesssim \Delta_0$ a typical experimental case.\cite{schoene2}\\
Since the coupling of the superconductor to the Kondo impurity is small and branch crossing and Andreev reflection above the gap are both processes of higher order we find that we may safely neglect their contributions. We can thus employ the transmission coefficients for the normal-superconductor QPC multiplied with the Kondo density of states $\Gamma_K^2/\left[(\omega - V)^2 + \Gamma_K^2\right]$ for single electron transmission and $\Gamma_K^4/\left\{\left[(\omega - V)^2 + \Gamma_K^2\right]\left[(\omega + V)^2 + \Gamma_K^2\right]\right\}$ for Andreev reflections to obtain
\begin{eqnarray*}
T_{eK}(\omega) &=& \frac{4\Gamma_{e}}{(1+ \Gamma_{e})^2} \frac{\Gamma_K^2}{(\omega - V)^2 + \Gamma_K^2},\nonumber\\
T_{AK}(\omega) &=& \frac{4\Gamma_{A}^2}{\Gamma_{A}^4+ 2 \Gamma_A^2(1- \Gamma_e^2) + (1+ \Gamma_e^2)^2} \nonumber\\
&& \times \frac{\Gamma_K^4}{\left[(\omega - V)^2 + \Gamma_K^2\right]\left[(\omega + V)^2 + \Gamma_K^2\right]} .
\end{eqnarray*}
The total CGF, including the background DOS, is given by $\ln \chi_{KS} = \ln \chi_{KNQS} + \ln \chi_{b}$, where $\chi_{KNQS}$ can be obtained from $\ln \chi_{NS}$ in Eq. \ref{fcssn} by replacing $T_e$ by $T_{eK}$ and $T_A$ by $T_{AK}$ while setting $T_{A2} = T_{BC} = 0$.\\
We calculate the ratio of the differential conductances in the case of a KNQS junction and a KNQN device. Fig. \ref{fig4} shows the comparison of the theory outlined above with experimental data, where the Kondo conductance peak in the KNQN state has been approximated by a Lorentzian.\cite{0957-4484-15-7-056} We observe a fairly good agreement up to a small difference for small voltages. This can be attributed to additional diffusive transport channels.
\begin{figure}[ht]
\includegraphics[width=7cm]{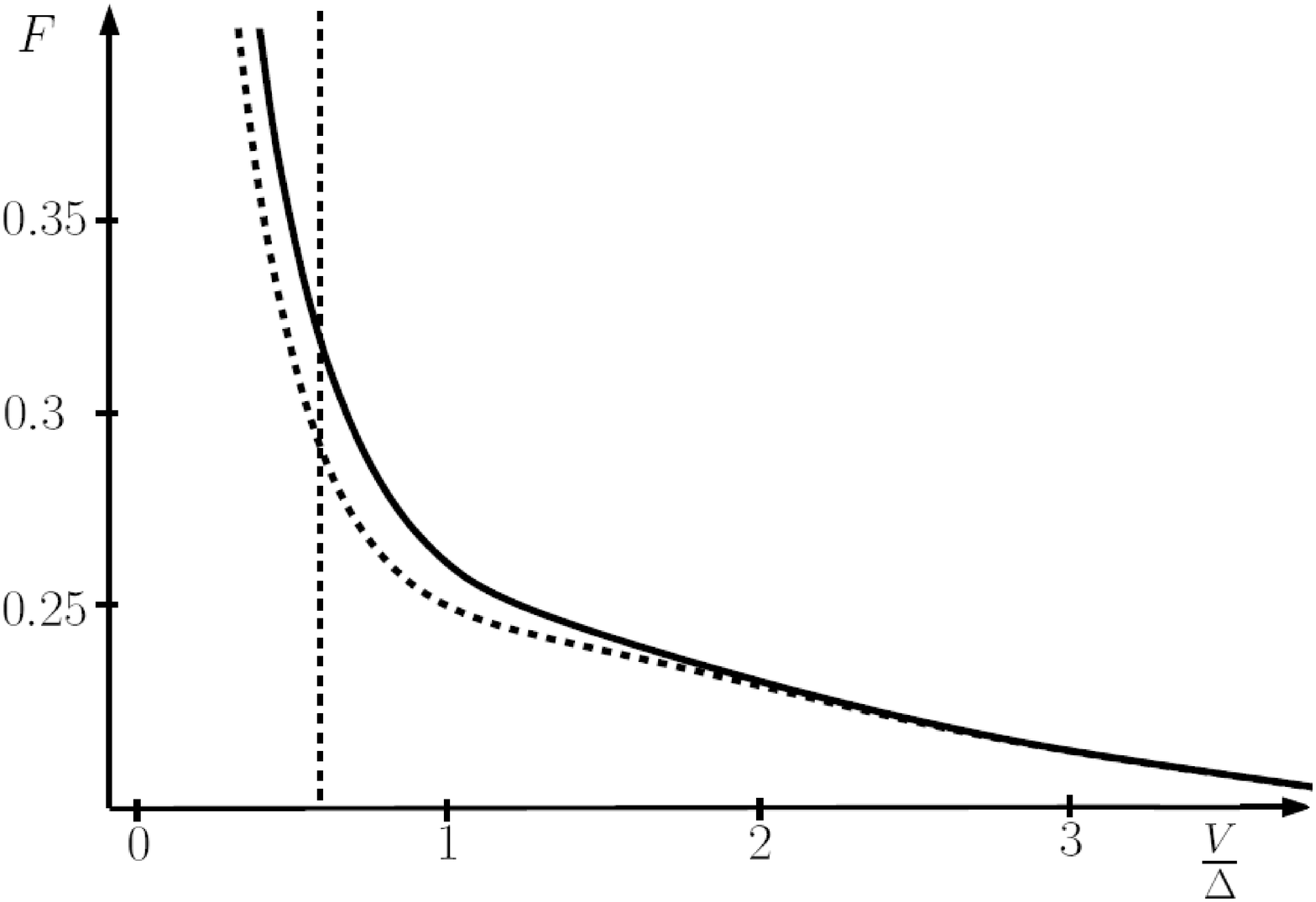}
\caption{Theoretical prediction for the Fano factor in the normal-quantum dot-superconductor junction with the same experimental parameters as used in the fit in Fig.~\ref{fig4}. The solid curve has been obtained with Andreev reflections and the dotted curve is the prediction without Andreev reflection ($T_{AK} = 0$). One observes a clear difference between the two curves that amounts up to a difference of about 15\% at $V/\Delta = 0.6$ (see dotted line).}
\label{fig5}
\end{figure}\\
In contrast to the prediction by the semiconductor model\cite{tinkham} we obtain quantitative agreement for voltages up to approximately $2\Delta$ due to the explicit inclusion of Andreev reflection and using the full energy dependence of the transmission coefficients derived in the Hamiltonian approach. The energy-dependent transmission coefficients lead to broader conductance peaks close to the superconductor gap while the explicit inclusion of Andreev reflection leads to an enhanced conductance at low bias. For higher voltages the structure in the background density of states starts to play a role and deviations are expected.
By introducing a transmission coefficient $T_{AK}$ we also included the possibility of Andreev reflection. Due to the low transparency of the contact it is strongly suppressed and causes only a small contribution to the differential conductance at low voltages. However, according to our effective theory they should be observable by a noise measurement.\\
In Fig.~\ref{fig5} the Fano factor including Andreev reflections has been compared to the prediction without Andreev reflections ($T_{AK} = 0$). One observes a clear difference in the Fano factors. This effect can be used to identify the presence of Andreev processes even for large interaction strengths.\\
The possibility of Andreev reflection also raises the question whether one could be able to observe a positive cross correlation in a three terminal setup like in [\onlinecite{2009arXiv0910.5558W,PhysRevLett.88.197001,PhysRevB.78.224515,soller}]. From the results obtained here we can address the case of a superconductor coupled to two Kondo dots contacted by two normal leads at the same chemical potential (see Fig. \ref{fig:scnn}).
\begin{figure}[ht]
\includegraphics[width=5cm]{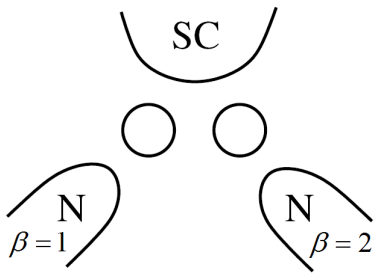}
\caption{Sketch of the experimental setup for measuring the cross-correlation. A superconductor (SC) is contacted via two quantum dots in the Kondo regime to two normal leads (N) at the same chemical potential.}
\label{fig:scnn}
\end{figure}\\
In principle a positive cross correlation of the currents in the two normal leads could be inferred from the process of crossed Andreev reflection where a Cooper pair from the superconductor splits up and the two electrons are transferred to the separate normal terminals. The current originating from this process has been analyzed in a similar geometry in [\onlinecite{PhysRevB.63.165314}]. For simplicity here we only want to address the case of $T=0$, where the CGF adopts the form
\begin{widetext}
\begin{eqnarray}
\ln \chi_{SNN}({\boldsymbol \lambda}, \tau) &=& 2 \tau \int_0^V \frac{d\omega}{\pi} \ln \{ 1+ \sum_{\beta = 1,2} T_{eK\beta} (\omega) (e^{i \lambda_\beta} -1)\} \theta \left(\frac{|\omega| - \Delta_0}{\Delta_0}\right) + \tau \int_{-V}^V \frac{d\omega}{\pi} \ln \{1+ \sum_{\beta = 1,2} T_{AK\beta} (e^{2i \lambda_\beta} -1) \nonumber\\
&& + T_{CAK} (e^{i (\lambda_1 + \lambda_2)} - 1)\} \theta \left(\frac{\Delta_0 - |\omega|}{\Delta_0}\right),
\end{eqnarray}
where the transmission coefficients can be inferred using the result from [\onlinecite{soller}] to be
\begin{eqnarray*}
T_{eK\beta} &=& \frac{4 \Gamma_e}{(1+ \Gamma_e)^2} \frac{\Gamma_{K\beta^2}}{(\omega - V)^2 + \Gamma_{K\beta}^2},\\
T_{AK\beta} &=& \frac{4 \Gamma_A^2}{\Gamma_{A}^4+ 2 \Gamma_A^2(1- \Gamma_e^2) + (1+ \Gamma_e^2)^2} \frac{\Gamma_{K\beta}^4}{[(\omega - V)^2 + \Gamma_{K1}^2] [(\omega + V)^2 + \Gamma_{K2}^2]},\\
T_{CAK} &=& \frac{8 \Gamma_A^2}{\Gamma_{A}^4+ 2 \Gamma_A^2(1- \Gamma_e^2) + (1+ \Gamma_e^2)^2}  \frac{\Gamma_{K1}^2 \Gamma_{K2}^2}{[(\omega - V)^2 + \Gamma_{K1}^2] [(\omega + V)^2 + \Gamma_{K2}^2]}.
\end{eqnarray*}
\end{widetext}
The cross correlation can be calculated as a second derivative of the CGF
\begin{eqnarray}
P_{12}^I = - \frac{1}{\tau} \left. \frac{\partial^2 \ln \chi_{SNN}({\boldsymbol \lambda}, \tau)}{\partial \lambda_1 \partial \lambda_2}\right|_{{\boldsymbol \lambda} = 0}.
\end{eqnarray}
\begin{figure}[ht]
\includegraphics[width=7cm]{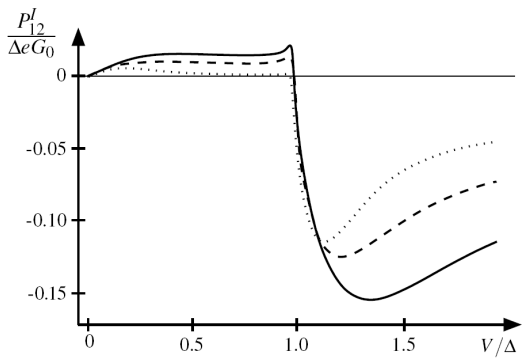}
\caption{Theoretical predicition for the cross correlation $P_{12}^I$ at $T=0$ in a three terminal setup with one superconducting and a normal lead. All curves have been obtained for $\Gamma= 0.05$ and a broadening of the BCS density of states has been taken into account by a Dynes parameter of $\Gamma_D = 0.005\Delta_0$.  The solid curve refers to $\Gamma_{K1} = 0.7\Delta_0 =\Gamma_{K2}$, the dashed curve is for $\Gamma_{K1} = 0.3\Delta_0, \; \Gamma_{K2} = \Delta_0$ and the dotted curve has been obtained for $\Gamma_{K1} = 0.3 \Delta_0 = \Gamma_{K2}$.}
\label{fig6}
\end{figure}\\
Fig. \ref{fig6} shows the result for three different configurations of the widths of the respective Kondo resonances. In all cases we observe a positive cross-correlation that first increases towards the superconducting gap and then quickly dissapears for voltages above the gap. This means that in principle it should be possible to observe a positive cross-correlation of the currents in the normal leads that is mediated by crossed Andreev reflection. Above the superconducting gap single electron transfer to the separate terminals sets in and causes a negative cross-correlation. However, we should mind that the prediction has been done for $T=0$. At finite temperature thermally activated single-electron transfer becomes possible. This process causes a negative correlation of the two currents that will, in most cases, destroy the positive cross-correlation by crossed Andreev reflection. Nevertheless, this shows that even in the Kondo regime a positive cross-correlation could in principle be observable.

\section{Conclusion}
We have calculated the FCS for a quantum dot in the Kondo regime coupled to a normal drain and a superconductor/STM tip. We have demonstrated the validity of our approach by comparing our results with existing studies as well as experimental data. For the normal-quantum dot-superconductor junction we found that according to our effective theory Andreev reflections could be seen in the noise, leading to a considerable change in the respective Fano factor.  Finally, we have investigated crossed Andreev reflection in a three-terminal setup. We have shown that a positive cross-correlation could in principle be observed even in the Kondo situation.

The authors would like to thank S. Maier and D. F. Urban for many interesting discussions. The financial support was provided by the DFG under grant No. 2235/3, and by the Kompetenznetz "Funktionelle Nanostrukturen III" of the Baden-W\"{u}rttemberg Stiftung (Germany).

\end{document}